\documentclass[sigconf]{acmart}
\AtBeginDocument{%
  \providecommand\BibTeX{{%
    \normalfont B\kern-0.5em{\scshape i\kern-0.25em b}\kern-0.8em\TeX}}}

\setcopyright{acmlicensed}
\copyrightyear{2024}
\acmYear{2024}
\acmDOI{XXXXXXX.XXXXXXX}

\copyrightyear{2024} 
\acmYear{2024} 
\setcopyright{acmlicensed}\acmConference[TREW @ CHI '24]{Trust and Reliance in Evolving Human-AI Workflows Workshop @ CHI Conference on Human Factors in Computing Systems}{May 11--16, 2024}{Honolulu, HI, USA}
\acmBooktitle{Trust and Reliance in Evolving Human-AI Workflows Workshop @ CHI Conference on Human Factors in Computing Systems, May 11--16, 2024, Honolulu, HI, USA}
\acmDOI{10.1145/XXX.XXX}
\acmISBN{979-8-4007-0330-0/24/05}

\begin{document}

\title[Just Like Me]{Just Like Me: The Role of Opinions and Personal Experiences in The Perception of Explanations in Subjective Decision-Making}

\author{Sharon Ferguson}
\email{sharon.ferguson@mail.utoronto.ca}
\affiliation{%
  \institution{Mechanical and Industrial Engineering, University of Toronto}
  \city{Toronto}
  \state{Ontario}
  \country{Canada}
}

\author{Paula Akemi Aoyagui}
\affiliation{%
  \institution{Faculty of Information, University of Toronto}
  \city{Toronto}
  \state{Ontario}
  \country{Canada}
}

\author{Young-Ho Kim}
\affiliation{%
  \institution{NAVER AI Lab}
  \country{Republic of Korea}
}

\author{Anastasia Kuzminykh}
\affiliation{%
  \institution{Faculty of Information, University of Toronto}
  \city{Toronto}
  \state{Ontario}
  \country{Canada}
}






\renewcommand{\shortauthors}{Ferguson et al.}

\begin{abstract}
As large language models (LLMs) advance to produce human-like arguments in some contexts, the number of settings applicable for human-AI collaboration broadens. Specifically, we focus on subjective decision-making, where a decision is contextual, open to interpretation, and based on one's beliefs and values. In such cases, having multiple arguments and perspectives might be particularly useful for the decision-maker. Using subtle sexism online as an understudied application of subjective decision-making, we suggest that LLM output could effectively provide diverse argumentation to enrich subjective human decision-making. To evaluate the applicability of this case, we conducted an interview study (N=20) where participants evaluated the perceived authorship, relevance, convincingness, and trustworthiness of human and AI-generated explanation-text, generated in response to instances of subtle sexism from the internet. In this workshop paper, we focus on one troubling trend in our results related to opinions and experiences displayed in LLM argumentation. We found that participants rated explanations that contained these characteristics as more convincing and trustworthy, particularly so when those opinions and experiences aligned with their own opinions and experiences. We describe our findings, discuss the troubling role that confirmation bias plays, and bring attention to the ethical challenges surrounding the AI generation of human-like experiences.
\end{abstract}




\received{1 March 2024}
\received[revised]{19 April 2024}
\received[accepted]{21 March 2024}

\maketitle

\section{Introduction}

Human-AI collaborative decision-making aims for a complementary performance \cite{Bansal2021, bansal2019beyond, philips2021human, donahue2022human, hemmer2021human}, where human and AI partners together achieve a better outcome than they would individually. In such ambiguous and open-to-interpretation scenarios where there is no ground truth, \citet{10.1145/3544549.3585727} have explored the use of Large Language Models (LLMs) to support human-decision makers by surfacing various viewpoints \cite{lai2021towards}. One such example of these ambiguous scenarios is in the domain of hate speech detection, particularly sexism, where subtle hate speech is more challenging to identify and remove automatically \cite{khurana2022hate}. Benokraitis \cite{Benokraitis1997} describes subtle sexism as a less visible form of discrimination that is based on gender and is oftentimes undetected, accepted as normal, or even considered to be benevolent. In fact, research has shown assessment of sexism can be highly subjective, depending on an individual's personal values, gender ideologies and, thus, is open to interpretation \cite{MITAMURA2017101}. Specifically in a social media context, for example, the same tweet can be considered sexist by one person, but not sexist by another \cite{hall2016they}. 
This ambiguity can be risky: while there is a risk of harm if hateful posts are spread through the internet \cite{womenzahra}, studies have also proven that hate speech filtering algorithms can unintentionally harm LGBTQ communities when mistakenly flagging words that would be considered offensive in other contexts \cite{oliva2021fighting, ramesh2022revisiting}. They also significantly over-restrict African American English \cite{davidson-etal-2019-racial}.  

Recent research suggests that human-AI collaboration, towards any goal, requires the AI not just to make a recommendation (i.e., to remove or not remove a potentially sexist post from social media) but also to be able to explain the reasoning behind it in a way that is \emph{relevant}, \emph{convincing} and \emph{trustworthy} for the user \cite{schaekermann2020ambiguity, wang2021explanations}. 
Further, when we specifically consider these subjective cases, \citet{alm2011subjective} argue that we need to move beyond traditional metrics, and evaluate user satisfaction, which can be measured in numerous ways. Firstly, \emph{relevance} is often used as an evaluation criterion for explanations  \cite{schaekermann2020ambiguity, kumar2021machine, hendricks2021generating}, largely based on how useful they are to the human-decision maker and how closely they represent the scenario or decision in question. Secondly, explanations must be \emph{convincing} to enable collaborative decisions, especially in ambiguous settings \cite{schaekermann2020ambiguity} and previous work \cite{palmer2023large} has shown LLM-produced arguments can be as persuasive as those human-authored ones. Then thirdly, \emph{trustworthiness} is pointed to as a key criterion for acceptance of a recommendation in decision-making tasks, with plenty of work exploring how Explainable AI (XAI) can be leveraged to build \cite{wang2021explanations, fuji2020trustworthy, kunkel2019let, hoffman2018metrics}, and calibrate trust \cite{zhang2020effect, zhou2019effects, wischnewski2023measuring, buccinca2021trust}. While truthfulness is undoubtedly also an important metric in objective decision-making, it is a challenging evaluation metric for subjective cases where there is no one ground truth.  



In summary, in many subjective decision-making scenarios, one's personal values and lived experiences can heavily influence one's assessment of a scenario. Given the importance of values, beliefs, and experiences, it would follow that the presence of beliefs or experiences in AI-generated input influences how a human perceives the input. While we know that current LLMs can generate uniquely human-like attributes, and humans find this troubling \cite{svenningsson2019artificial}, little is known about how users perceive personal opinion and experience representations in explanations when considered in subjective decision-making. In the context of politics, one study found that sharing personal experiences about harm was more convincing than sharing facts \cite{kubin2021personal}, as everyone could agree that harm should be avoided. Other work has shown that LLMs can generate arguments \cite{palmer2023large}, and that they commonly reiterate certain opinions \cite{santurkar2023whose}. However, we don't yet know how these findings expand to other contexts, and how humans evaluate them.  
Thus, we address the following research question: 1) Are opinions and experiences perceived by humans in AI-generated explanations? And, 2) How are these opinions and experiences perceived by users? 

To answer these research questions, we conducted an empirical study with 20 participants to explore how they evaluate human and AI-generated text explanations in the context of subtle sexism. We ask participants to imagine that they are in a decision-making context and have to evaluate whether the provided scenario is sexist or not using the explanation provided. We ask them to assess who authored the explanation, and how \emph{relevant}, \emph{convincing} or \emph{trustworthy} it is. We found that personal opinions and personal experiences were identified in both human- and AI-authored explanations, and participants described these as \emph{trustworthy}. Further, an alignment between the opinions and experiences shown in the explanation and the participant's own beliefs exaggerated this effect, suggesting harmful cognitive biases at play. We hope to bring this scoped finding to the CHI community to start a discussion on how the human-like ability to generate personal beliefs and experiences influences perceptions of trust, and how we can consider this finding in the design of collaborative systems. 



\section{Method}

\begin{figure*}[h]
\includegraphics[width=0.65\textwidth]{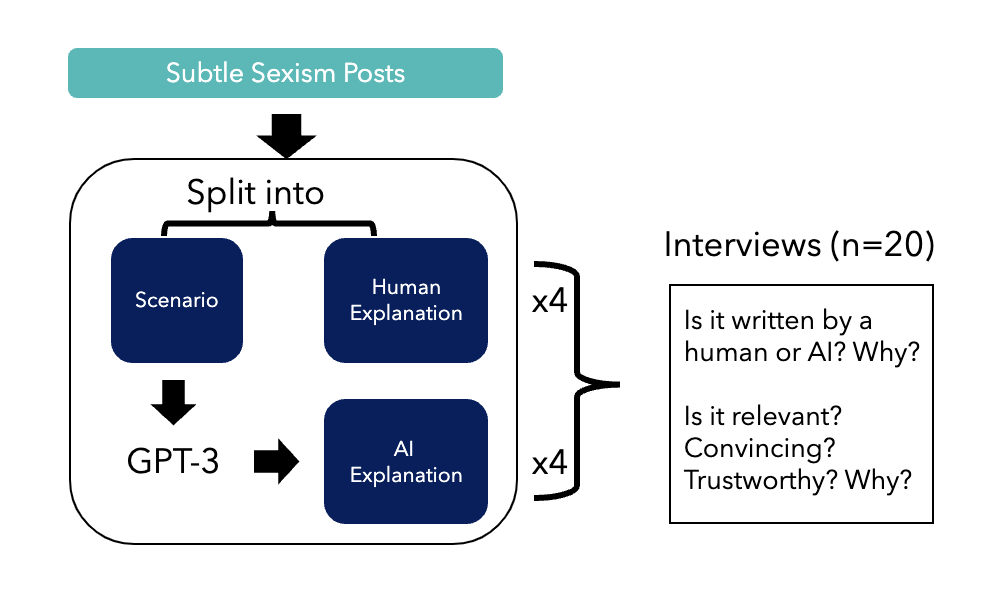}
\caption{Overview of the research methodology}
\label{Fig1}
\end{figure*}

We conducted a set of semi-structured interviews with 20 participants to gauge their perception of human and AI-generated explanations. This study was approved by the university's research ethics board, and all participants provided informed consent. Participants were shown eight scenarios with accompanying explanations of subtle sexism, and asked to imagine themselves as part of collaborative decision-making on whether the scenario constitutes a case of subtle sexism. Our research methodology can be seen in Figure \ref{Fig1}. 

The scenarios and human explanations, were collected from online discussion sites such as Reddit\footnote[1]{\url{www.reddit.com}}, The Everyday Sexism Project\footnote[2]{\url{www.everydaysexism.com}}, and Twitter\footnote[3]{\url{www.twitter.com}}. We selected scenarios (descriptions of events) that were paired with an interpretation or explanation of the scenario and why it is or is not sexist. The complete dataset contained 117 scenarios and accompanying explanations, which are representative of the ``everyday explanations'' that humans use when discussing sexism. To collect the AI-generated explanation text for these scenarios, we used GPT-3~\cite{DBLP:journals/corr/abs-2005-14165}, which was the state-of-the-art large language model at the time. We prompted the model using the question-answer feature, asking ``\texttt{Is this scenario sexist: \{\{scenario\}\}. Why or why not?}'' To ensure that a coherent explanation-text was generated for each scenario, we prompted the model three times per scenario, resulting in 351 AI-generated explanation texts. 

To keep the interviews at an appropriate length, we chose eight scenario and explanation pairs to present to the participants --- four explanations which were generated by GPT-3, and four which were collected from online discussion sites. From the larger dataset, we chose eight scenario and explanation pairs based on the following criteria: the explanation was coherent; there was a balance of argumentative stance (it is sexist vs. it is not sexist); and the length was appropriate for an interview, with both the scenario and explanation being less than five sentences. These chosen explanations also represented some of the higher-quality human explanations in the dataset. The text-based output from LLMs can be displayed to users in multiple modalities, which we know influence the perception of explanations \cite{robbemond2022understanding}. As part of the larger study, we manipulated whether users were presented the explanation in text or audio form, though we do not focus on the outcome of this manipulation in this short paper. 

In the semi-structured interviews, we collected demographic information and asked introductory questions to gauge participants' familiarity with AI technology. Of the 20 participants, 10 identified as women, nine as men and one as non-binary. Participants averaged 30 years old (min: 20, max: 56) and spanned various roles from student to company executive both within and outside of AI. Most participants said they often use conversational AI, while few used chatbots regularly. The rest of the interview contained eight scenario and explanation pairs, each with the same line of questioning. We started this portion of the study by explaining the collaborative decision-making context, and asking participants to imagine that they had to decide on whether a given scenario was sexist or not, and they had input from another party, who could be a human or an AI, but they were not aware of which one. We also briefly described how the AI-generated texts were produced by stating that it was not a model specifically trained on sexism, but just a general language model.  We showed participants the scenario and explanation, and asked whether the explanation was generated by a human or an AI model, and why. We then asked participants to rate and explain their rating for the explanation on three qualities: \emph{relevance}, \emph{convincingness}, and \emph{trustworthiness}, based on the dictionary definitions for these terms. At the end of the interview, if participants were interested, we shared which explanations were human and AI-generated.
Literature shows \emph{relevance}, \emph{convincing} and \emph{trustworthy} can be evaluated objectively and subjectively \cite{wang2021explanations, schaekermann2020ambiguity, hendricks2021generating, herlocker2000explaining, buccinca2020proxy, hoffman2018metrics}. Thus, we collected both quantitative (scales) and qualitative data for eight explanations across 20 participants. To protect the privacy of the posters whose scenarios and explanations are used in this study, we provide descriptions for the collected scenarios and explanations in Table \ref{tab:my-table}, but do not provide the verbatim text. Verbatim text is provided for AI-generated content.

We anonymized and transcribed interviews before following the Braun and Clarke thematic analysis method \cite{clarke2015thematic}. Two researchers went through two rounds of initial coding, where three randomly chosen interviews were open-coded individually by both researchers each time. The researchers met after each coding round to consolidate and organize the open codes, resulting in two iterations of the coding scheme before it was finalized. At this time, the researchers re-coded all twenty interviews using this finalized coding scheme. While we also collected quantitative data for the \emph{relevant}, \emph{convincing}, and \emph{trustworthy} scales, in this paper we focus on the qualitative results. 

\begin{table*}[t]
\footnotesize
\centering
\caption{Description of the Scenarios and Explanations used in the interview study. Verbatim scenarios and explanations (written by humans) are not included to protect poster privacy, though AI-generated text is included verbatim in italics. H=Human-authored, AI = AI-authored.}
\label{tab:my-table}
\begin{tabular}{|p{0.4cm}| p{7cm}| p{0.4cm} |p{7cm}|}
\hline
\# & Scenario        &\#                                                                                                                                                                                                                                                                                                                                                                                                                                                                                                                                         & Explanation                                                                                                                                                                                                                                                                                                                                                                                        \\ \hline
1     &  The scenario contains commentary on the ``massively disproportionate'' number of women taking STEM subjects in school, and how this may be caused by ``society's attitude towards women and these areas''. They discuss people providing surprised comments when women say they are studying high-level science. & H1 

 & The explanation states that while these comments are not meant to offend, they are often made at critical times when children are choosing a career path, and can thus cause women to stay out of science to prevent them from being seen as ``abnormal''.
 
 \\\hline
2     & The scenario where an adult women is referred to as a girl, by a man, in a workplace setting.    & H2                                                                                                                                                                                                                                                                                                                                                                                                     & The explanation states that this scenario drives the reader ``insane'' and it would not be appropriate to do the same to a man.                                                                                                                                                                           \\ \hline
3     & The scenario just contains the word ``hysterical.''           & H3                                                                                                                                                                                                                                                                                                                                                                                                                                                                                                                        & The explanation describes how the word is triggering because the Greek root of the word refers to the word for ``uterus'', and the word is rarely used to describe a man.
                                               \\ \hline
4     & The scenario describes a family where the father is upset if the daughter does not help to cook or clean, but the brothers are allowed to relax.       & H4                                                                                                                                                                                                                                                                                                                                                                                                  & The explanation states that while people talk about ``men's work'', we should not assign chores based on gender. They describe their experience enjoying yard work growing up, and how some men would prefer to be in the kitchen. They say both ways are sexist.                                                                                                         \\ \hline
5    &  The scenario describes many examples of men boiling thing down to a woman's ``biology''. The examples include being emotional due to your period, being able to handle pain due to childbirth, and women getting drunk faster than men.        & AI1                                                                                                              & \textit{It is sexist. It's a way of undermining women by reducing their capabilities as human beings to nothing more than their physical features. It's a way of saying that women have no control over their behavior. It's insulting. And it's not helping. }                                                                                                                                                                                                                                                                                                                    \\ \hline
6    & The scenario contains two quotes from books containing the lines ``courage uncommon in a woman'' and ``grip that was decidedly unfeminine.''     & AI2                                                                                                                                                                                                                                                                                                                                                                                     & \textit{Yes, and I think that's sexist because it is saying that a woman not acting like a woman is admirable.        }                                                                                                                                                                                                                                                                                                                                          \\ \hline
7& The scenario describes when the term ``guys'' is used when addressing both men and women.     & AI3                                                                                                                                                                                                                                                                                                                                                                                                                                                                        & \textit{No, it isn't sexist. The word 'guys' is not a gendered word. It is not sexist.             }                                                                                                                                                                                                                                                                                                                                                                                                                                                                                                            \\ \hline
8   &  The scenario describes ``benevolent sexism'' and gives the examples of women being called ``dear'' by senior male colleagues or ``girls'' when hanging out with other female co-workers.  & AI4                                                                                                                                                                                                                                              & \textit{I am not sure that it is sexist. It is a term that refers to the way men can try to be inclusive of women or supportive of women but with condescending or patronising undertones. I think it is important to look at how we treat each other and to look at how we might be sexist without meaning it.                       }                                                                                                           \\ \hline
\end{tabular}%

\end{table*}

\section{Results}

In this section, we provide evidence for the perception of both personal opinions and experiences within human and AI explanation text. 

\subsection{Recognizing Opinions and Experiences}







Participants recognized the elements of personal opinion and personal experience in explanations, both those authored by humans and AI-generated. Personal opinion was simply identified 20 times in our interviews
--- mostly in reference to both actual and perceived human-authored explanations. Personal opinions were defined as when the explanation revealed personal beliefs or points of view, for example:

\begin{quote}
    ``\textit{...So it had a very strong opinion and then it supported the opinion afterwards by again, just kind of distilling down what the scenario was talking about...}'' [P4]
\end{quote}

Personal experiences were identified 56 times in the interviews, 
and were primarily discussed in terms of both actual and perceived human examples. Personal experiences refer to the presence of a personal example or the way something personally affected the explanation author, such as:

\begin{quote}
    ``\textit{This sounds like it comes from a place of having experienced this}''. [P5]
\end{quote}

We found that these explanation elements were identified by participants across scenarios, thus being scenario independent: opinions and experiences were recognized by participants in reference to seven of eight scenarios. We also found no evidence that individual differences between participants drove this identification: opinions were identified by 11 out of 20 participants, and experiences by 17.



Both opinion and experience were often brought up when describing why a participant believed the explanation was written by a human (note that this does not mean the explanation was actually human-generated):

\begin{quote}
    ``\textit{It's coming from a very personal story perspective. So it's one person's perspective which is good. So I wouldn't say it's a trustworthy news source, but at the same time, I do trust it as someone's own personal opinion}'' [P12]
\end{quote}

Participants described how the specific examples contained in explanations had to come from lived experiences, and would be challenging to train an AI to replicate: 

\begin{quote}
   ``\textit{...It's actually...coming up with real-life circumstances and scenarios of why this might not be sexist...So I feel like whoever explained this has real-world experience and it's not just something that was trained to say the right thing}". [P5]
\end{quote}

Even going as far as to say that it would be unnerving to think of an AI that could generate text as if from personal experience: 

\begin{quote}
    ``\textit{...it was kind of reflecting on their own domestic chore experiences and bringing that into the argument that just instantly made me feel like it was a human cause you don't want to think about an AI like that. It's just a bit unnerving...}'' [P9]
\end{quote}



In terms of evaluation of the explanation, participants noted that personal opinions are worth considering:




\begin{quote}
    ``\textit{Trustworthy? Yeah. I mean it is based on someone's opinion and it is based on a different situation that the speaker had provided. So is it reliable? Yes, it's reliable.}'' [P8]. 
\end{quote}



One participant even shared that personal experiences are only convincing when they are real -- meaning, they reflect a real event that a human experienced, suggesting that an AI imitation of this would not be convincing:

\begin{quote}
``\textit{But I think if I was standing there and I was in a conversation and there was a human woman saying this in rebuttal, and I would say it's a four or a five because it's someone who is speaking from lived experience.}'' [P15]
\end{quote}

However, many also noted that sometimes an author's opinion or description of their experience is not enough to make an explanation \emph{trustworthy} and \emph{convincing}. For that effect to be achieved, the opinion or experience must be backed by facts, supporting evidence or sources.

\begin{quote}
    ``\textit{Okay, I don't see any sort of source for these facts that are being stated. I don't see any, yeah so it's just an opinion of AI or a person and I would have to see evidence.}'' [P2]
\end{quote}



Some participants argued that a \emph{trustworthy} explanation would contain both personal components such as opinions and experiences, as well as logic, facts, or statistics:

\begin{quote}
``\textit{I think yes [it is trustworthy], because it seems like it doesn't come only from academic experience but also from personal experience firsthand. So it seems, yeah I think it appeals to our senses to trust one that can handle both [facts and opinions].}'' [P10]
\end{quote}

In summary, participants recognize the presence of opinions and experiences displayed in human- and AI-generated explanations, across various scenarios. The presence of these explanation elements made explanations more \emph{convincing} and \emph{trustworthy} when used in subjective decision-making, though they should be combined with objective facts and sources.

\subsection{Comparing Opinions and Experiences}

Interestingly, even though we did not directly inquire about the participant's opinions about a scenario or their similar personal experiences, they often offered these as a justification for their answers. In fact, they recognized the author's opinion in the explanation and discussed how this opinion did or did not align with their personal opinion, and how this alignment, or lack thereof, influenced their assessment of the explanation. For example:

\begin{quote}
``\textit{...it has a lot of truths to it. You don't really hear someone use the word hysterical to describe another man.}'' [P7]
\end{quote}

This comparison with their own opinion was discussed 97 times, in reference to all scenarios, by 19 participants, and overwhelmingly in response to perceived human explanations. Despite the fact that humans often made this comparison in regard to explanations they thought were human-authored, these explanations were often actually AI-authored. 

This comparison echoes literature \cite{MITAMURA2017101} that exposed the weight of personal values when considering sexism. Overall, when an opinion displayed in the explanation aligned with the participant's opinion, they were more likely to assess a human wrote it. When an explanation aligned, or related, to their personal experience in the world, participants described that this felt ``human'':

\begin{quote}
   ``\textit{This feels like a response that I would have, personally. This is something probably that I would see myself saying. So I would guess that this is a human response.}'' [P4]
\end{quote}

\begin{quote}
``\textit{I find it relatable, that explanation to, to my career, to my job and everything. So I feel connected to that explanation or that makes me feel that it was done by a human.}'' [P11]
\end{quote}

In terms of comparing experiences displayed in the explanation to their own experiences, this was found 48 times in our interviews, in regard to all scenarios and brought up by 17 participants. In this case, participants made this comparison in explanations that both were actually and perceived to be human-authored. 

While in general, we found that opinions and experiences had a more important role in the evaluation of \emph{convincing} and \emph{trustworthy}, rather than \emph{relevant}, there were notable examples where participants mentioned alignment with personal experience in their assessment of \emph{relevance}. Perhaps they used their personal experience as a middle ground between the scenario and explanation. If the scenario aligned with their personal experience, as did the explanation, the explanation was relevant to the scenario: 

\begin{quote}
``\textit{I think it's definitely relevant, and even I believe that the word hysterical is typically associated with women and that kind of reinforces this misconception that we as a society have that women are the ones who super emotional and they get carried away and they can act in a crazy manner or be hysterical. So yeah, I think the explanation makes sense to me}'' [P13]
\end{quote}

Furthermore, when participants noticed alignment with their personal opinions and experiences within the explanation, they would find the explanation to be more \emph{Convincing}:

\begin{quote}
    ``\textit {I like the reasoning, it's pretty similar... It's just in line with my personal values system}" [P3]
\end{quote}


\begin{quote}
``\textit{I did relate to it. I don't see hysterical referred to, [or] used as a descriptor for men or haven't historically seen that. And so that's the piece where it was like, oh yeah, that is an observed behavior that I've also noticed. So I feel convinced by that...}''[P15]
\end{quote}

Further, a lack of alignment makes things less \emph{convincing}:

\begin{quote}
    \textit{``...if I was a guy and you're trying to convince me based on this explanation, I wouldn't really be convinced because that's what I'm used to hearing the entire time. I'm used to saying that's what I, that's the way everyone around me talks. Yeah. So I wouldn't be convinced by this explanation.''} [P15]
\end{quote}

In terms of \emph{trustworthiness}, participants described an emotional connection, or an emotional appeal, that was brought about when the explanation aligned with their experience: 

\begin{quote}

    \textit{''And I think because it also speaks to some of my own experiences and the experiences of some of my friends and colleagues growing up, I'm, it just intuitively fosters this connection...[it] speaks about an experience that a lot of people have had growing up and choosing what they wanna do in life and their career paths. So there is some emotional appeal that is going on there''} [P9]
\end{quote}


    


In addition, a few participants also considered that while they agree with the explanation's stance, someone else with a different opinion might make a different assessment:

\begin{quote}

\textit{``Again, same thing. I do agree with what's been said, so I'm like, yeah, I trust that a lot. But if I put [myself] in someone else's shoes, if I did not agree with what's been said would it be trustworthy? I think it's a three. I could react or someone could react back and say, `Ah, I have a different opinion'. Of course, go ahead. So based on what I know and what I believe, I trust it a lot...}" [P8]
    
\end{quote}

And lastly, we found that in some instances, participants said that the lack of alignment with their own opinion made it hard to evaluate the explanation. In this case, the participant recognized the impact that this lack of alignment had on their evaluation, and was thus unable to provide a rating:


\begin{quote}
``\textit{I don't really have enough information to make the decision if it's convincing or trustworthy mainly because I don't agree with the explanation. So I'm not really finding an answer to whether it's trustworthy or convincing.}'' [P17]
\end{quote} 

In summary, participants tended to automatically compare the opinions and experiences they perceived within the explanation to their own opinions and experiences. Whether or not these opinions and experiences aligned with their own influenced their evaluation of the explanation, making it more \emph{convincing} and \emph{trustworthy}, and evening making it challenging to evaluate the explanation if there is a lack of alignment.

\section{Discussion and Implications}

In this work, we argue that as LLM's abilities advance, they are becoming more suitable collaboration partners for humans, specifically in the context of subjective decision-making. The role of AI output in these contexts is to present new information and perspectives to the human decision-maker. Explanations in collaborative, subjective decision-making are less likely to be grounded in facts than explanations for objective decisions; it is more common that opinions and personal experiences comprise these new perspectives. As opinions and personal experiences may be considered uniquely human attributes, we were interested in whether and how these attributes were perceived in AI explanations. We found that humans did identify opinions and experiences in explanations for subtle sexism scenarios, and perceived them to be both \emph{convincing} and \emph{trustworthy}, making them important for subjective decision-making. 
While an argumentative stance (i.e., sexist or not sexist) can be depicted in some non-textual explanation formats, it would be hard to share personal opinions (such as ``I don't think women should be confined to traditionally feminine interests'') and personal experiences, which are normally described in a narrative format, in forms other than text. Thus, word-based explanations are perceived to contain opinions and experiences which are important for subjective decision-making.

Personal opinions and experiences were also often attributed to human authorship. As has also been shown in past work in the context of emotions \cite{10.1145/3490100.3516464}, we have demonstrated that these modern LLMs can generate plausibly human text elements; in our case, opinions and experiences. Thus, while these elements may aid in subjective decision-making, we have to consider the ethical implications of participants potentially believing these explanations come from humans. In fact, we provide evidence that the relationship between the explanation author and the explanation evaluation is causal in some cases --- some participants described how they value human's individual experiences, and thus they would trust a human's explanation of their experience, but not an AI-generated replication. This means that if we use LLMs in subjective decision-making collaborative systems, we must be sure to disclose their contributions as AI-generated, even if this might harm the \emph{trustworthiness} of the explanation. In many contexts, humans distrust AI-generated text \cite{10.1145/3290605.3300469, 10.1145/3441000.3441074}, thus future work can investigate which elements or factors need to be present in AI-generated text to calibrate trust.

Explanations which featured opinions and experiences were also found to be \emph{convincing} and \emph{trustworthy}, which aligns with past work in other contexts, where researchers found that personal experiences regarding politics are more convincing than facts \cite{kubin2021personal}, and make people seem more rational and worthy of respect \cite{scheve2022willingness}. However, opinions alone were often noted as not enough to completely convince participants. This finding suggests one way in which we may need to fine-tune language models if used in this context. Our results also show that personal opinions and experiences, in conjunction with statistics or other forms of evidence, would be most convincing --- as was also suggested in past work on political disagreement \cite{kubin2021personal}. Future work can identify how to prompt or adjust the design of language models to provide this balance.

Perhaps the more troublesome finding was that participants unpromptedly judged how these opinions and experiences aligned with their own, greatly influencing their overall perception. When an explanation contains an opinion or experience similar to the participant's own, they tend to rate the explanation overall as more \emph{trustworthy} and \emph{convincing}. This is known as a confirmation bias --defined as ``seeking or interpreting evidence in ways that are partial to existing beliefs'' \cite[p. 175]{nickerson1998confirmation} -- 
which poses the risk of reinforcing the user's existing opinions and negating the intention of providing new perspectives to the decision-making process. Thus, we have to be careful when deciding on the opinions and experiences present in these explanations. Perhaps in these collaborative decision-making settings, we should present participants with multiple LLM outputs, representing multiple opinions and experiences, or even prompt models to describe different perspectives. It has been shown that these models can provide different opinions, although they may provide one more commonly than another \cite{santurkar2023whose}. Further, recent work has shown that LLMs can be trained to generate widely accepted outputs that can help people with diverse viewpoints reach consensus \cite{bakker2022fine}, suggesting that these models could be used both to generate different perspectives and resolve them. This would have the benefit of being perceived as \emph{convincing} and \emph{trustworthy}, but still providing new perspectives and information that can help participants make their decisions.

\section{Limitations}
Due to our collection of subtle sexism scenarios from naturally occurring internet discussion sites, we limit the variation of scenarios studied. For instance, we found that the large majority of the posts regarding subtle sexism argued for why a scenario was sexist, limiting our ability to assess human explanations arguing why the scenario was not sexist. Further, because we could only fit a small number of scenario-explanation pairs in the interview, we also cannot comment on the generalizability of our findings to subtle sexism in various contexts and other types of hate speech.

Another limitation of this work is the small sample size. The majority of our study participants (though not all) came from Western communities, and may have different beliefs and lived experiences from those in other parts of the world. Additionally, because online forums and social media tend to be populated mostly by Western communities, this could influence the alignment that we see between experiences displayed in explanations and the participant's own experiences. Future work should extend this investigation to understand how cultural differences may influence the perception of alignment. Furthermore, LLMs have been shown to reflect societal biases, and can create hate-speech themselves. While we filter out this type of content in our study, future work can measure the biases in these LLM-generated texts, and how biased opinions and experiences influence human perceptions.

Lastly, as part of the larger scope of the study, we asked participants to describe why they believe a text to be human or AI-authored. This focus prevented us from being able to analyze how knowledge of the explanation source affected perception, which is an important next step in the work. 

\section{Conclusion \& Future Work}

In this work, we present an interview study of participants' perceptions of human and AI explanations for subjective decision-making, specifically using the example of identifying subtle sexism. We argue that in subjective cases, multiple perspectives presented in collaboration can be helpful. To make this feasible, we motivate this work with the idea that LLMs can be used to generate these perspectives. We ask participants to evaluate human and AI-generated explanations as if they were participating in this collaborative process. We found that participants often perceived the explanations, both those authored by humans and AI, to contain personal opinions and experiences. The presence of these elements typically leads participants to view the explanation as \emph{convincing} and \emph{trustworthy} and also believe that it was written by a human. Further, we show that whether these opinions and experiences are aligned with the participant's own opinions and experiences is even more important for trust, highlighting a troubling tendency to conform to confirmation bias, negating the original intent of collaborative decision-making. Thus, we show that these elements of explanations are particularly important for collaboration in subjective hate-speech detection, and we motivate future work to address how we might best provide multiple, differing opinions and experiences to collaborative human decision-makers, and how we can avoid ethical challenges that arise when AI generates human-like opinions and experiences.

\bibliographystyle{ACM-Reference-Format}


\end{document}